\documentclass[twocolumn,superscriptaddress,floatfix,preprintnumbers,amssymb ,amsmath]{revtex4-1}
\usepackage{graphicx,amssymb}
\usepackage{color,ulem}\usepackage{bm}  
\usepackage{float}

\renewcommand{\Re}{\mathop \mathrm{Re}}

\begin{document}
\title{Optimized proximity thermometer for ultra-sensitive detection}
\author{Bayan Karimi}
\affiliation{QTF Centre of Excellence, Department of Applied Physics, Aalto University School of Science, P.O. Box 13500, 00076 Aalto, Finland}
\author{Danilo Nikoli\'c}
\affiliation{Fachbereich Physik, Universit\" at Konstanz, D-78467, Germany}
\author{Tuomas Tuukkanen}
\affiliation{QTF Centre of Excellence, Department of Applied Physics, Aalto University School of Science, P.O. Box 13500, 00076 Aalto, Finland}
\author{Joonas T. Peltonen}
\affiliation{QTF Centre of Excellence, Department of Applied Physics, Aalto University School of Science, P.O. Box 13500, 00076 Aalto, Finland}
\author{Wolfgang Belzig}
\affiliation{Fachbereich Physik, Universit\" at Konstanz, D-78467, Germany}
\author{Jukka P. Pekola}
\affiliation{QTF Centre of Excellence, Department of Applied Physics, Aalto University School of Science, P.O. Box 13500, 00076 Aalto, Finland}
\affiliation{Moscow Institute of Physics and Technology, 141700 Dolgoprudny, Russia}
\date{\today}

\begin{abstract}
We present a set of experiments to optimize the performance of the noninvasive thermometer based on proximity superconductivity. Current through a standard tunnel junction between an aluminum superconductor and a copper electrode is controlled by the strength of the proximity induced to this normal metal, which in turn is determined by the position of a direct superconducting contact from the tunnel junction. Several devices with different distances were tested. We develop a theoretical model based on Usadel equations and dynamic Coulomb blockade which reproduces the measured results and yields a tool to calibrate the thermometer and to optimize it further in future experiments.
\end{abstract}


\maketitle
\section{Introduction}
Virtually any parameter depending on temperature, preferably monotonically, can form a basis for thermometry~\cite{OVL, Quinn}. Yet depending on the application, one needs to make a choice of the system and technique based on several criteria, including sensitivity, noise, power dissipation, physical size, and speed of the thermometer. Besides these criteria, one often needs to consider whether the measured quantity can be obtained theoretically from a well-known, preferably simple physical law without fit parameters: if this is the case the technique may qualify as "primary thermometry". However, most of the time, like in the present work, this is not the case, and we deal with "secondary thermometry". To measure local temperature of nanostructures at very low temperatures ($< 1$ K) have been recently developed with several techniques~\cite{Mavalankar, Maradan, ZBA, Libin, JP2, Nicola, Borzenets, Du}. Here we build on a technique based on temperature dependent proximity superconductivity yielding sensitive thermometry with ultra-low dissipation. The technique is particularly well adaptable to calorimetric detection of tiny heat currents as well as fast thermometry towards the lowest temperatures in mesoscopic systems on-chip. The main goal of the present investigation is to optimize the sensitivity (responsivity) of the sensor and to model its behaviour using well established theoretical framework. The main results of the current work are (i) one order of magnitude increased sensitivity of the device with respect to the earlier realization~\cite{ZBA} and (ii) a full theoretical account of its characteristics.

\begin{figure}
	\centering
	\includegraphics [width=\columnwidth] {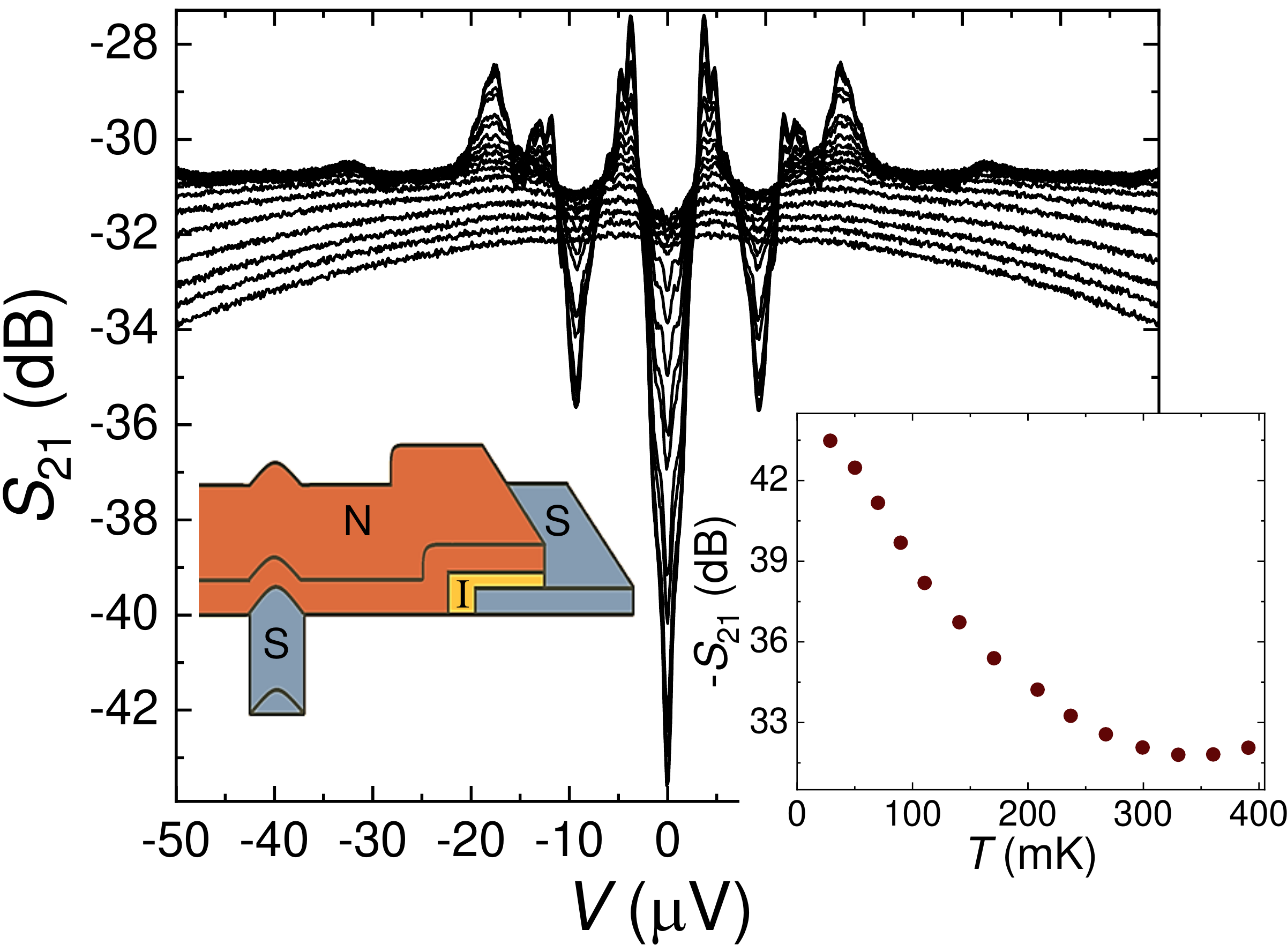}
	\caption{Principle and basic characteristics of the thermometer. The left inset: schematic illustration of the thermometer tunnel junction. Here N, S, and I stand for normal metal, superconductor, and insulating barrier, respectively. The main panel shows the rf transmission $S_{21}$ of the proximitized junction at various temperatures from $\sim 20$~mK up to $\sim 400$~mK, at $-140$~dBm applied power. This signal is directly related to the conductance of the junction. Right inset: zero-bias conductance ($-S_{21}$) as a function of bath temperature taken from the data in the main panel. 
		\label{fig1}}
\end{figure}
\section{Description of the thermometer}
The thermometer that we study is schematically shown in the left inset of Fig. \ref{fig1}. The normal lead of a standard NIS (normal metal-insulator-superconductor) junction is connected to another superconducting lead via direct metal-to-metal contact. This lead induces the proximity effect to the N lead and permits supercurrent via the tunnel junction. The basic characterization of this thermometer is presented in Ref. \cite{ZBA}.

This thermometer has been recently operated in a set-up that allows one to monitor temperature and its variations on $\mu$s time scales \cite{BJNET}. The main panel of Fig. \ref{fig1} presents the dc bias voltage $V$ dependent rf transmission $S_{21}$ at the resonance of the $LC$ circuit loaded by the thermometer junction in parallel, measured at various temperatures. In this case, the superconducting contact is at the distance of $L=450$ nm from the tunnel junction. For low conductance $dI/dV$ of the junction, the variations of $S_{21}$ are proportional to $-dI/dV$. In the figure we thus observe temperature dependent conductance of the thermometer junction. The favourable operation of the thermometer is at $V=0$, where the temperature dependence of $S_{21}$ on $T$ is strongest, and the self-heating $IV$ is minimal.

The right inset of Fig. \ref{fig1} depicts the temperature dependence of $-S_{21}$ of this junction measured at $V=0$, using a very small excitation power (-140 dBm) to measure. We see that $-S_{21}$ presents an almost linear increase with decreasing temperature well below 300 mK, thus providing a sensitive and non-invasive thermometer. These characteristics are to be compared to the temperature dependent dc conductance results that will be presented below.   

In the current work we limit ourselves to detecting the conductance and current voltage characteristics of the junction in a quasi-dc measurement. The samples were fabricated on a commercially available silicon wafer onto which a $300$~nm layer of silicon oxide has been grown. In order to have stable tunnel junctions we used a $22$~nm suspended germanium hard mask. Moreover all different kinds of samples were made on the same chip in one fabrication process, meaning that this process for all of the them was equal and the dominant difference between them was their geometry. We use electron-beam lithography for writing the patterns on the chip and three-angle electron-beam evaporation of the metal films. All samples have $35$~nm copper as normal metal and $20$~nm as both superconducting Al leads. The insulator in the tunnel junction is a thin layer of aluminum-oxide formed by letting pure oxygen into the chamber on top of one of the aluminum films. The thermometers were made in a single vacuum cycle allowing fabrication of clean metallic contacts without additional cleaning of the samples.

\begin{figure}
	\centering
	\includegraphics [width=\columnwidth] {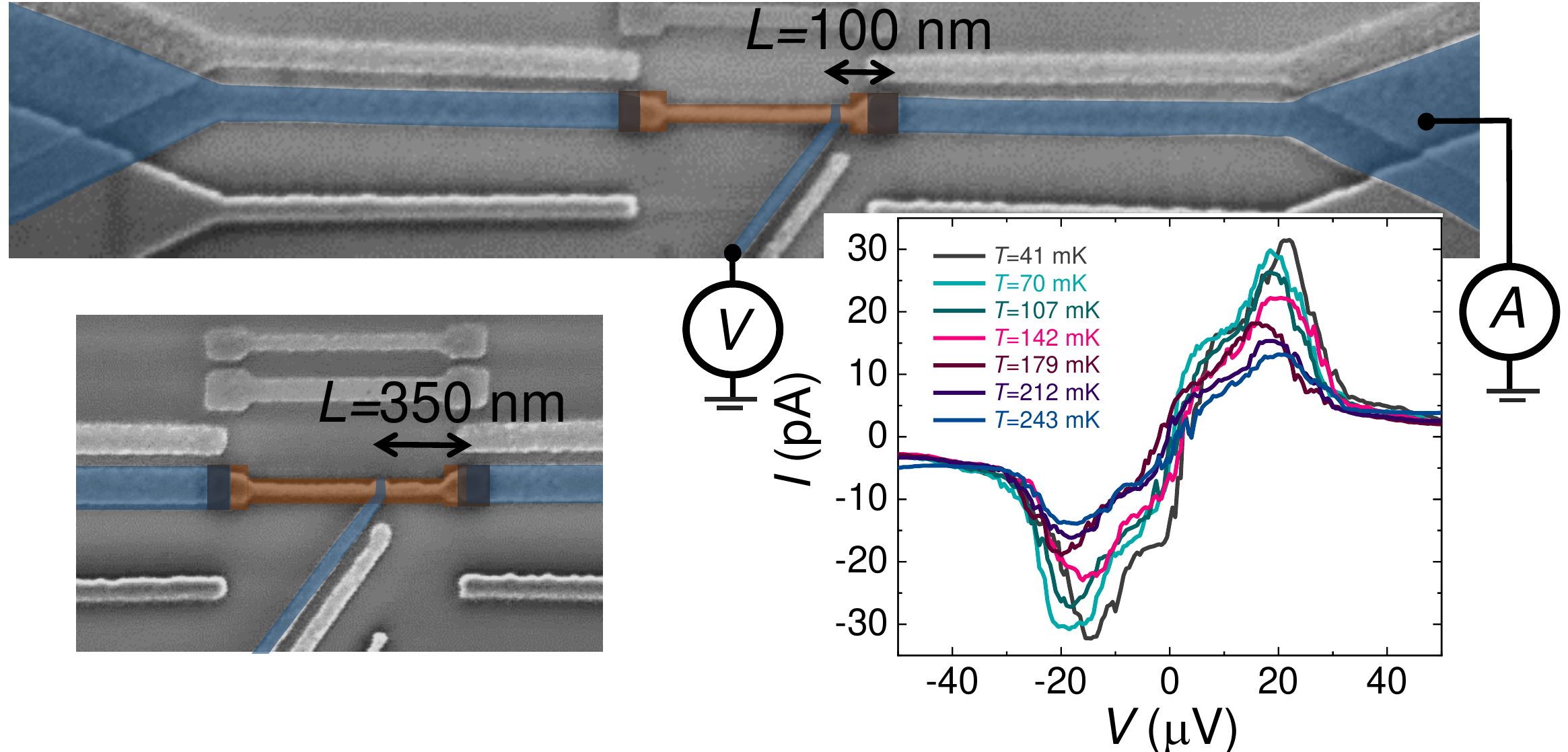}
	\caption{Scanning electron micrographs (SEM) of two different samples. The normal metal (brown) is coupled to three superconducting leads (blue): two (left and right) via tunnel barriers (gray) and one via a direct contact. The distance $L$ between the right junction with clean contact varries from $50$~nm to $350$~nm in $50$~nm intervals (SEM images of two of them $L=100$~nm in the main frame and $L=350$~nm at left bottom). We present in this paper data on transport between the right and middle contact. (Right bottom) The IV characteristics of the sample with $L=100$~nm at different bath temperatures.
		\label{fig2}}
\end{figure}
\section{Experimental results and optimization of the sensitivity}
An important figure of merit of a sensor is its responsivity, which for this thermometer reads $\mathcal{R}=|dS_{21}/dT|$. The apparent noise in a temperature measurement is then inversely proportional to $\mathcal{R}$ as long as noise is not intrinsic originating from true temperature fluctuations. Intuitively the responsivity is expected to increase when the proximity is enhanced, by bringing the clean contact closer to the junction. Therefore we fabricated several proximity junctions with nominally equal parameters, apart from the differing distance $L$. Figure \ref{fig2} shows the scanning electron microscope (SEM) images of two samples ($L=100$~nm in the main panel and $L=350$~nm in the left-bottom). Along with the two SEM images, it shows in the right inset the measured IV characteristics of sample with $L=100$~nm at different bath temperatures (40-240 mK) which indicates the non-vanishing current in the small bias range with maxima at $\pm 20~\mu$V, due to the induced proximity effect in the normal-metal island. It is clear that the peak current $I_{\rm max}$ decreases due to the decrease of proximity effect by increasing the bath temperature.  
\begin{figure*}[t!]
	\begin{center}
	\includegraphics [width=18 cm] {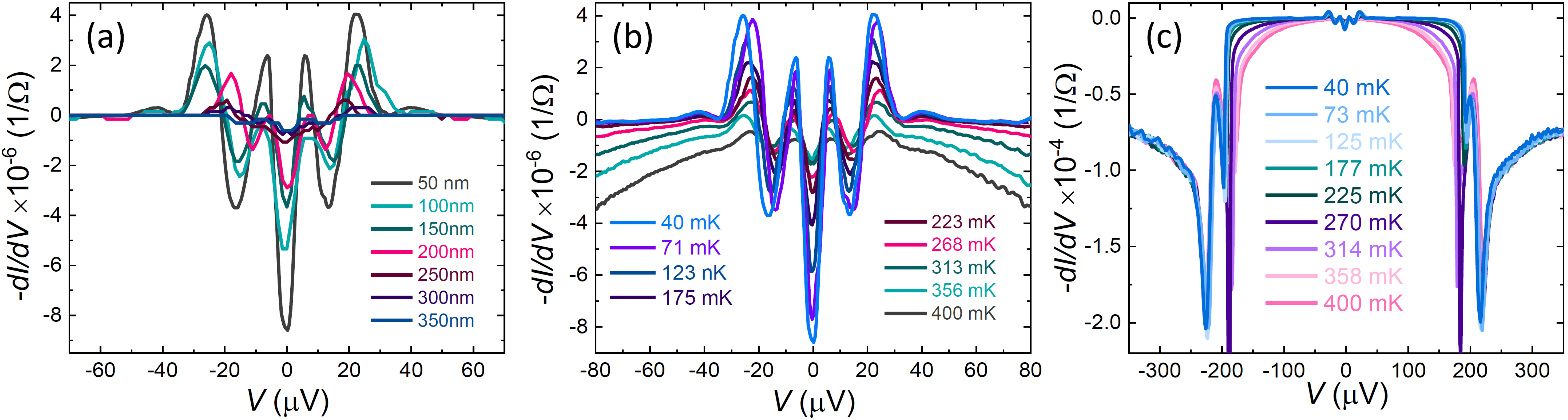}
	\end{center}
	\caption{Bias $V$ dependence of the conductance $G=dI/dV$ of the thermometers under different experimental conditions. (a) $-dI/dV$ for various samples with $L=50-350$~nm at temperature $T=40$~mK. (b) A similar plot as (a) but just for one thermometer with $L=50$~nm at various temperatures from $T=40$~mK up to $400$~mK. (c) As in (b) but now in a wider bias range demonstrating the onset of quasiparticle current at $V\simeq \pm \Delta_0/e=200~\mu$V besides the zero-bias anomalies.
		\label{fig3}}
\end{figure*}
We measured the conductance $dI/dV$ of the junctions with low frequency $\sim 10$~Hz Lock-in techniques. The measured conductance of the proximitized junction as a function of applied voltage bias under different conditions is shown in Fig. \ref{fig3}. In panel (a) this dependence is shown for different distances $L=50-350$~nm at $50$~nm intervals and fixed bath temperature $T=40$~mK. A few reproducible features can be observed at sub-gap regime within $V\sim \pm 50~{\rm \mu V}\ll \Delta_0/e$, where $\Delta_0$ is the superconducting gap. The sharpest one of them is our favorable feature at zero-bias voltage, i.e., zero bias anomaly (ZBA). All these features get suppressed by increasing the distance $L$. Figure \ref{fig3}b demonstrates the main feature, the temperature dependence of conductance for $L=50$~nm i.e. the thermometer with the strongest ZBA feature. Sensitivity of zero-bias conductance down to the lowest temperature is obvious. The overall change of the baseline and nearly parabolic bias dependence of it are due to quasiparticle current arising at finite temperatures. Figure \ref{fig3}c is a zoom-out of \ref{fig3}b that emphasizes conductance character due to quasiparticle current at voltages around the superconducting gap, $V\simeq \pm 200~\mu$V. The standard BCS coherence peaks at $eV=\pm \Delta_0$ of an NIS junction are now split due to the existance of the minigap in the proximitized normal metal.
\section{Modeling and comparison to the experiment}
\begin{figure}[tb]
	\centering
	\includegraphics[scale=0.33]{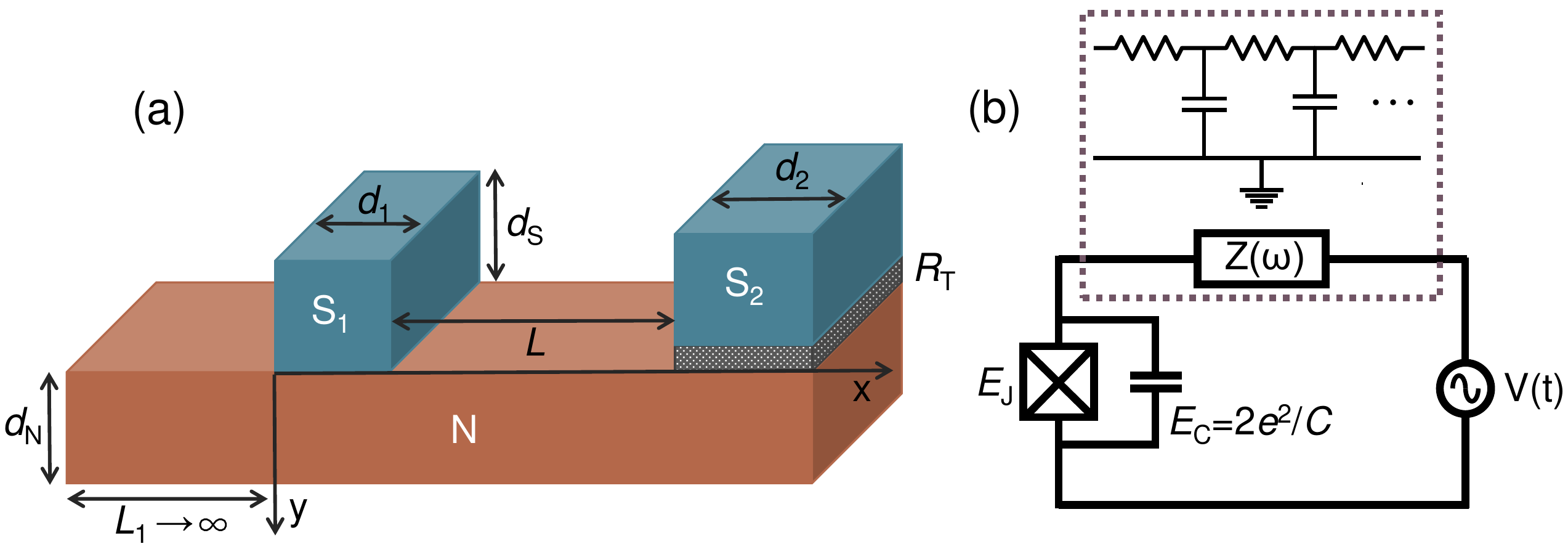}
	\caption{(color online) (a) Scheme of the overlap $\rm SNIS$ junction where two superconductors $\rm S_1$ and $\rm S_2$ (blue) are set on top of a semi-infinite normal metal wire (orange). $\rm S_1N$ is a clean contact of a length $d_1$, while $\rm S_2N$ is a tunnel contact of a length $d_2$ and a resistance $R_T$. Thicknesses of the superconductors and the normal metal are $d_{\rm S}$ and $d_{\rm N}$, respectively. (b) Schematic representation of the circuit under study where the Josephson junction of a capacitance $C$ and a resistance $R_T$ is coupled to an infinite RC transmission line (see inset) of the impedance $Z(\omega)=\sqrt{R_0/i\omega C_0}$, where $R_0$ and $C_0$ are the resistance and the capacitance per unit length of the line. }
	\label{fig:SNIS}
\end{figure}
In this section we theoretically analyse an overlap SNIS junction, which is close to the experimental setup studied above, coupled to electromagnetic (EM) environment. Let us first discuss the SNIS junction with the geometry shown in Fig. \ref{fig:SNIS} (a) and described in the corresponding caption. All metallic parts are assumed to be in the dirty limit, in which the elastic mean free path $l_e$ is much smaller than the superconducting coherence length $\xi\sim\sqrt{\hbar D/2\Delta_0}$, where $D=v_{\rm F}l_e/3$ with the Fermi velocity $ v_{\rm F}$. In order to describe such a system we make use of the imaginary time quasiclassical Green's function formalism determined by the Usadel equation of motion \cite{Usadel1970,Belzig1999, Belzig1996}. If the total thickness of the superconductor and the normal metal of the system depicted in Fig.~\ref{fig:SNIS} (a) is sufficiently small $(d_{\rm N}+d_{\rm S} < \xi)$ we can neglect all the derivatives in the $y$ direction and average the Usadel equation over the width reducing it to an effectively one-dimensional problem described by the following ordinary differential equation \cite{Belzig1999}
\begin{equation}
\label{eqn:Usadel_theta}
\frac{\hbar D}{2}\frac{d^2\theta_n}{dx^2} = \omega_n\sin[\theta_n(x)] -\Delta(x)\cos[\theta_n(x)]\,.
\end{equation}
Here, $\theta_n(x)$ is the proximity angle of the normal metal, $D$ is the diffusion coefficient of the material, $\omega_n=(2n+1)\pi k_{\rm B}T$ are the fermionic Matsubara frequencies with the temperature $ T $ and $ n=0,\pm 1,\pm 2,\ldots $ and $\Delta(x)$ is the superconducting order parameter defined as follows
\begin{equation}
\Delta(x)=\frac{d_S}{d_{\rm N}+d_{\rm S}} \Delta
\end{equation}
for $0<x<d_1$ and $=0$ otherwise, where $\Delta$ is the order parameter of superconductor $S_1$. We note that the normal metal underneath $S_1$ effectively acts like a superconductor with the reduced superconducting gap, $d_{\rm S}/(d_{\rm S}+d_{\rm N})\Delta$.
To obtain a general solution, Eq.~(\ref{eqn:Usadel_theta}) has to be supplemented by the appropriate boundary conditions at the ends of the normal metal wire $\theta_n(L_1)=0$ and $\partial_x\theta_n|_{x=d_1+L+d_2}=0$. The other boundary conditions come from the continuity of the proximity angle function as well as the current conservation throughout the system, $\theta_n(a-0)=\theta_n(a+0)$ and  $\partial_x\theta_n|_{x=a-0}=\partial_x\theta_n|_{x=a+0}$ for $a=0,~d_1$ 
(see Fig.~\ref{fig:SNIS} (a))  \cite{Belzig1999,KuprianovLukichev1988}. Since ${\rm NS_2}$ is a tunnel contact with low transparency we neglect the proximity effect in this region. The boundary condition problem described above can be solved numerically by employing the finite difference method where Eq.~\eqref{eqn:Usadel_theta} is rewritten as a system of nonlinear algebraic equations. The normal and the anomalous Green's function components in $\theta$ representation read $G_{\omega_n}(x)=\cos[\theta_n(x)]$ and $F_{\omega_n}(x)=\sin[\theta_n(x)]$, respectively \cite{Belzig1999}.

Based on the solution of Eq.~(\ref{eqn:Usadel_theta}) for the overlap junction depicted in Fig.~\ref{fig:SNIS} (a) the critical current through a tunnel $\rm NS_2$ interface depends on the anomalous component of the Green's function and, therefore, is given by \cite{AmbegaokarBaratoff1963,Werthamer1966,LarkinOvchinikov1967}
\begin{equation}
\label{eqn:Ic}
I_c = \frac{\pi k_{\rm B}T}{2eR_T}\sum_{\omega_n>0}
F_{\omega_n}^S\bar{F}_{\omega_n}^N,
\end{equation}
where $R_T$ is the resistance of the tunnel junction, $T$ is the temperature and $F^S_{\omega_n}=\Delta(T)/\sqrt{\omega_n^2+\Delta(T)^2}$ is the anomalous Green's function of superconductor $S_2$. The temperature dependence of the  superconducting gap is assumed to be $\Delta(T)=\Delta_0\tanh(1.74\sqrt{T_c/T-1})$, with $T_c$ as the critical temperature of the superconductor \cite{Gap}. The proximity angle, $\theta_n(x)$, depends in the general case on the $x$ coordinate which means one is supposed to average the solution along the $\rm NS_2$ interface of a finite length $d_2$ obtaining $\bar{F}^N_{\omega_n}$ as follows 
\begin{equation}
\bar{F}^N_{\omega_n}=\frac{1}{d_2}\int_{d_1+L}^{d_1+L+d_2}\sin[\theta_n(x)]dx.
\end{equation}

Let us now discuss the contribution from the EM environment schematically represented by the circuit depicted in Fig.~\ref{fig:SNIS} (b) and described in the corresponding caption. Due to the dynamical Coulomb blockade the current mediated by the tunnelling of a Cooper pair in an ultrasmall Josephson junction of a capacitance $C$ is described by the so-called $P(E)$ function \cite{Averin1990, IngoldNazarov1992}
\begin{equation}
I_s(V)=\frac{\pi e E^2_J}{\hbar}\left[P(2eV)-P(-2eV)\right].
\end{equation}
Here $E_J=\hbar I_c/2e$ is the Josephson energy of the junction. The $P(E)$ function is the probability for an electron to emit a photon to the environment and it is defined as
\begin{equation}
\label{eqn:P(E)}
P(E)=\frac{1}{2\pi\hbar}\int_{-\infty}^{\infty}dt \exp\bigg[4J(t)+\frac{i}{\hbar}Et\bigg],
\end{equation}
where $J(t)=\langle[{\varphi}(t)-{\varphi}(0)]{\varphi}(0)\rangle$
is the equilibrium correlation function of the phase $\varphi(t)=(e/\hbar)\int_{-\infty}^t V(t')dt'$ of the voltage across the junction. This function depends on the total impedance of the system, $Z_t(\omega)$, as follows

\begin{figure}
	\centering
	\includegraphics [width=\columnwidth ] {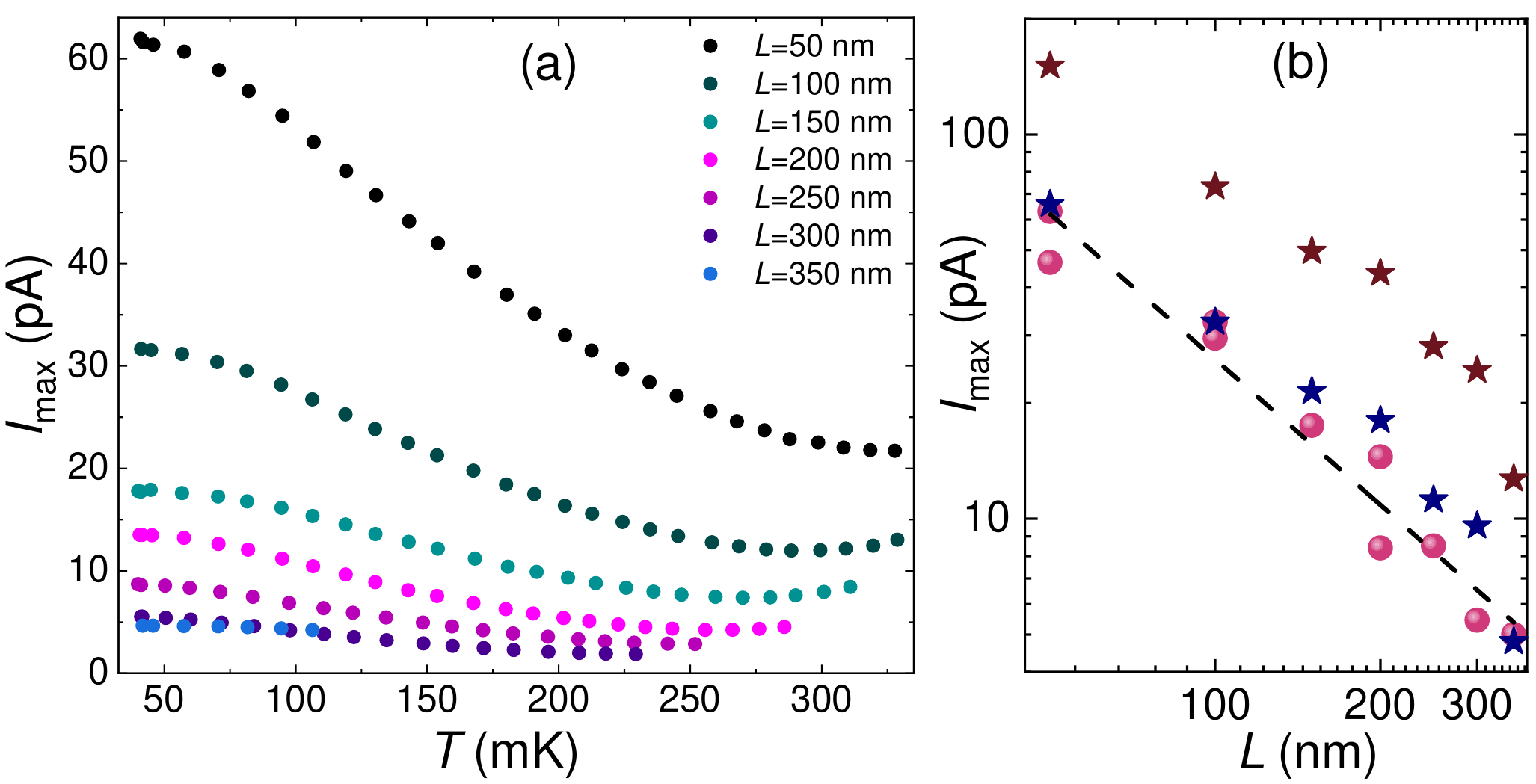}
	\caption{Peak current $I_{\rm max}$ of different measured samples. (a) Temperature dependence of $I_{\rm max}$ for junctions with varying $L$. (b) Base temperature values of $I_{\rm max}$. The pink dots are extracted from the measurement in (a). Star symbols are from theory whereas the brown ones are calculated at $T=40~$mK, while the blue ones are those with actual overheated temperatures as explained in the text. Dashed line $I_{\rm max}=7.9\times 10^3 L{\rm [nm]}^{-1.25}$~pA is a fit through experimental data. 
		\label{IV:T}}
\end{figure}
\begin{eqnarray}
J(t)=&&2\int_0^\infty \frac{d\omega}{\omega}\frac{\Re[Z_t(\omega)]}{R_K}\\&&\times
\bigg\{\coth\bigg(\frac{\hbar\omega}{2k_{\rm B}T}\bigg)[\cos(\omega t)-1]-i\sin(\omega t)\bigg\}\nonumber.
\end{eqnarray}
Here $R_K=h/e^2$ denotes the resistance quantum and $T$ is the environmental temperature. The total impedance of the system reads
\begin{equation}
Z_t(\omega)=\frac{1}{i\omega C+Z^{-1}(\omega)},
\end{equation}
where $C$ is the capacitance of the junction and $Z(\omega)$ is the impedance of the EM environment. In our model the EM environment is assumed to be an infinite RC transmission line whose impedance is $Z(\omega)=\sqrt{R_0/i\omega C_0}$, where $R_0$ and $C_0$ are the resistance and the capacitance per unit length of the line, respectively (see Fig.~\ref{fig:SNIS} (b)). Since the impedance of the RC transmission line depends on the ratio between $R_0$ and $C_0$, the appropriate
dimensionless parameter which characterizes the line is
$k=(R_0C/C_0)/R_K$~\cite{IngoldNazarov1992,Nazarov1989}.

\begin{figure}
	\centering
	\includegraphics[width=\columnwidth]{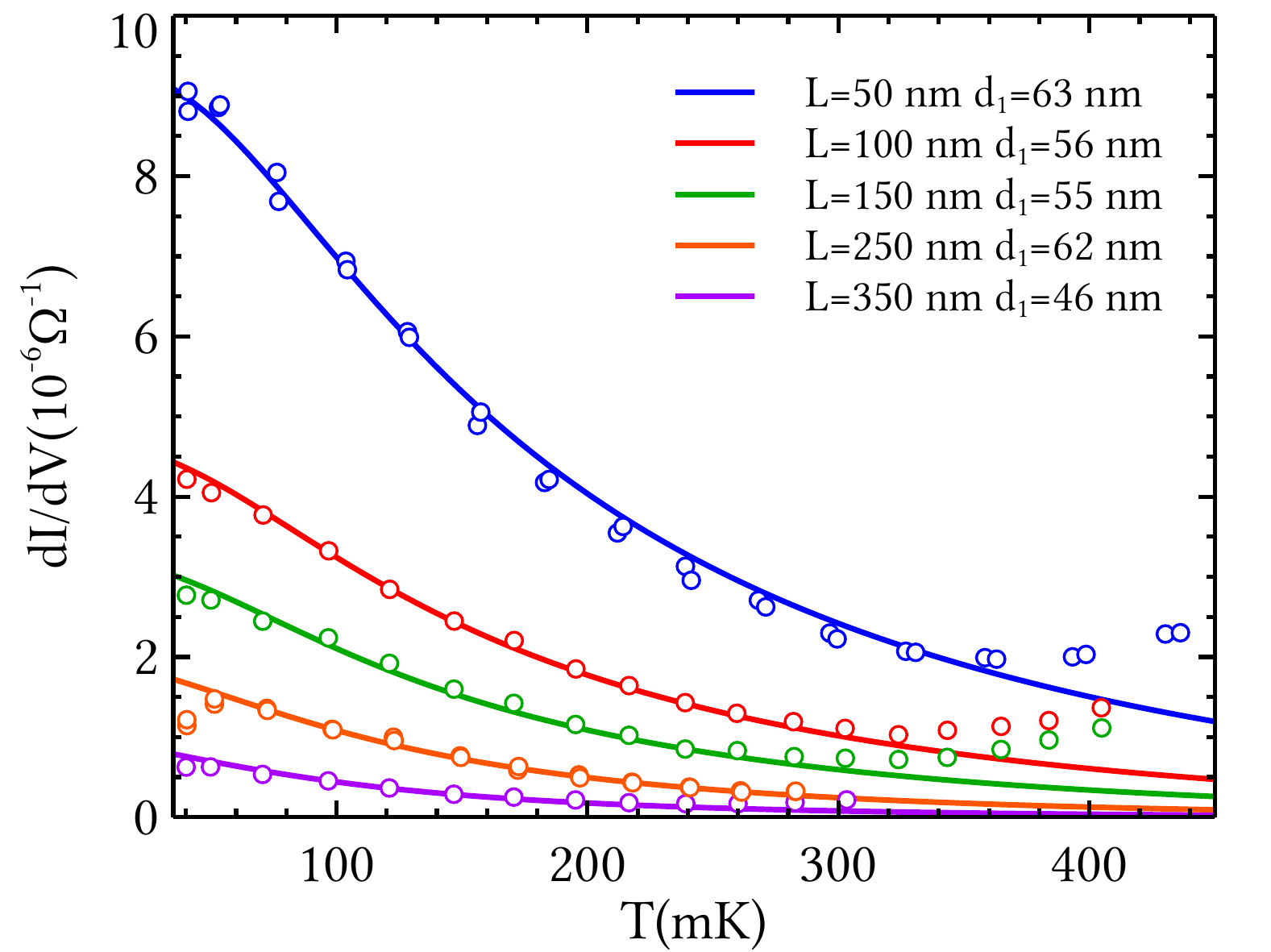}
	\caption{(color online) Comparison between measurement (circles) and theory (solid lines) for zero-bias conductance as a function of temperature. For the theory of junctions with various normal metal lengths $L$, we assume the charging energy $E_C=3.5\Delta_0=900 \mu$eV, $d_s/(d_N+d_s)=0.36$, $d_2=120$~nm, $d_1\in[46~{\rm nm}, 63~{\rm nm}]$, and $R_T\in[15~{\rm k}\Omega, 16.1~{\rm k}\Omega]$. The junction is assumed to be coupled to an infinite RC transmission line characterized by $k=0.0115$ (see the text for the definition of $k$).}
	\label{fig:ZBC}
\end{figure}
\section{Analysis and discussion of the data}
The measured IV characteristics like the ones shown in Fig. \ref{fig2}, allow us to extract the temperature and length dependences of different samples as shown in Fig. \ref{IV:T}. Panel (a) demonstrates the $T$ dependence for seven different samples. The main feature of all these data sets is the increase of $I_{\rm max}$ towards lower $T$ in accordance with the prediction of the theory. Yet, one can observe the saturation of $I_{\rm max}$ both at high and low $T$. At high $T$, this is because of the emergence of thermal quasiparticle current. More interestingly the current saturates below $\sim 100$~mK especially for samples with short $L$, a feature to be discussed below. In panel (b) we extract $I_{\rm max}$ for different samples at base $T\simeq 40$~mK. In Fig. \ref{IV:T}b we also include the theoretically calculated $I_{\rm max}$, the maximum of $I_s$, according to the theory presented in the previous section. Assuming $T=40$~mK in the calculation overestimates the $I_{\rm max}$ by about factor of three (brown stars in Fig. \ref{IV:T}b). This is very natural based of overheating of the proximitized normal-metal lead at finite bias voltage of about $V_{\rm max}=20~\mu$V, which is the position where the current is maximized. Quantitatively, writing the heat balance equation $I_{\rm max}V_{\rm max}=\Sigma {\mathcal{V}}(T^5-T_0^5)$, where $\Sigma=2\times 10^9~{\rm WK^{-5}m^{-3}}$ is the electron-phonon coupling constant of copper, $\mathcal{V}=1\times 10^{-21}~{\rm m^3}$ is the volume of the copper island and $T_0=40$~mK is the bath (phonon) temperature allows us to determine the temperature for each thermometer at this bias point. We obtain $T=125-215$~mK for samples with $L=350-50$~nm with $50$~nm intervals, respectively. Repeating the calculation of $I_{\rm max}$ at these temperatures for the corresponding samples, we obtain a much better agreement with the measured values of $I_{\rm max}$ as shown by the blue star symbols in Fig. \ref{IV:T}b. The message of this result and analysis is that it is very important to perform a true zero-bias measurement to avoid overheating. Applying even the tiny bias leads to severe self-heating of the thermometer.

Figure~\ref{fig:ZBC} shows a comparison between measured (circles) and theoretically predicted (solid lines) zero-bias conductance in the tunnel contact for various lengths of the junction. We obtain an excellent match. The measured zero-bias conductance in this figure shares similar temperature dependence with $I_{\rm max}$ in Fig.~\ref{IV:T}. There are important properties worth discussing in this data. First, the overall responsivity $\mathcal{R}$ of the thermometer improves by decreasing the length $L$ by one order of magnitude when $L$ shrinks from $350$~nm to $50$~nm. Second, unlike $I_{\rm max}$ in Fig.~\ref{IV:T} the responsivity is not lost even at base temperature; instead the dependence remains more or less linear in $T$. It is important to mention that the zero-bias conductance is not obtained for exactly $V=0$. It shows the averaged slope of the IV curves close to the zero voltage due to the experimental reasons as the $dI/dV$ is measured with finite voltage amplitude. 

\section{Conclusions}
We have found experimentally that the sensitivity of the SNIS thermometer operated at zero bias voltage can be enhanced dramatically by bringing the S contact to the very proximity of the tunnel junction, this way increasing the current through it. Specifically, we demonstrate that the zero-bias conductance measurement outperforms a standard IV measurement by avoiding self-heating at low temperatures. We have developed a theoretical model based on proximity superconductivity and dynamical Coulomb blockade, which captures quantitatively the measured data in its validity range. With this optimization, we have increased the responsivity of this thermometer by about one order of magnitude compared to the initial realization of the concept, and made it suitable for continuous detection of microwave quanta in GHz range~\cite{BJNET, Roope, Lee}.

This work was funded through the European Union's Horizon 2020 research and innovation programme under Marie Sklodowska-Curie actions (grant agreement 766025) and Academy of Finland grant 312057. D. N. thanks Aalto University for the hospitality during his visit. We acknowledge the facilities and technical support of Otaniemi research infrastructure for Micro and Nanotechnologies (OtaNano).

\end{document}